\documentclass[useAMS,usegraphicx,usenatbib]{mn2e}

\title[Properties of Starless and Prestellar Cores in Taurus]{Properties of Starless and Prestellar Cores in Taurus Revealed by {\it Herschel\/}\thanks
{{\it Herschel\/} is an ESA space observatory with science instruments provided
by European-led Principal Investigator consortia and with important 
participation from NASA.} SPIRE/PACS Imaging}
\author[K. A. Marsh et al.]{K. A. Marsh$^{1}$\thanks{E-mail:
Ken.Marsh@astro.cf.ac.uk}, 
M. J. Griffin$^{1}$, 
P. Palmeirim$^{2}$, 
Ph. Andr\'e$^{2}$, 
J. Kirk$^{3}$, 
\newauthor D. Stamatellos$^{3}$, 
D. Ward-Thompson$^{3}$, 
A. Roy$^{2}$,
S. Bontemps$^{4,5}$,
J. Di Francesco$^{6}$,
\newauthor D. Elia$^{7}$,
T. Hill$^{8}$,
V. K\"onyves$^{2,9}$,
F. Motte$^{2}$,
Q. Nguyen-Luong$^{10}$,
N. Peretto$^{1}$,
\newauthor S. Pezzuto$^{7}$,
A. Rivera-Ingraham$^{11,12}$,
N. Schneider$^{4,5}$,
L. Spinoglio$^{7}$,
and G. White$^{13,14}$ \\
$^{1}$School of Physics and Astronomy, Cardiff University, Cardiff CF24 3AA, UK\\
$^{2}$Laboratoire AIM, CEA/DSM-CNRS-Universit\'e Paris Diderot, IRFU /
Service d'Astrophysique, C.E. Saclay, Orme des Merisiers, \\
91191 Gif-sur-Yvette, France \\
$^{3}$Jeremiah Horrocks Institute for Astrophysics and Supercomputing,
University of Central Lancashire, Preston, PR1 2HE, UK\\
$^{4}$Univ. Bordeaux, LAB, UMR 5804, 33270, Floirac, France \\
$^{5}$CNRS, LAB, UMR 5804, 33270, Floirac, France \\
$^{6}$National Research Council of Canada, Hertzberg Institute of
Astrophysics, 5071 West Saanich Rd., Victoria, BC, V9E 2E7, Canada \\
$^{7}$Istituto di Astrofisica e Planetologia Spaziali - INAF, Via Fosso
del Cavaliere 100, I-00133 Roma Italy \\
$^{8}$Joint ALMA Observatory, Alonso de Cordova, 3107, Vitacura, Santiago,
Chile \\
$^{9}$Institut d'Astrophysique Spatiale, UMR8617, CNRS/Universit\'e
Paris-Sud 11,91405 Orsay, France \\
$^{10}$Department of Astronomy \& Astrophysics, University of Toronto,
50 George Street, Toronto, ON M5S 3H4, Canada \\
$^{11}$Universit\'e de Toulouse, UPS-OMP, IRAP, Toulouse, France \\
$^{12}$CNRS, IRAP, 9 Av. Colonel Roche, BP 44346, F-31028 Toulouse Cedex 4,
France \\
$^{13}$Department of Physics and Astronomy, The Open University, Walton Hall,
Milton Keynes, MK7 6AA, UK \\
$^{14}$RALSpace, Rutherford Appleton Laboratory, Chilton, Didcot OX11 0NL, UK}
\begin{document}

\pagerange{\pageref{firstpage}--\pageref{lastpage}} \pubyear{2002}

\maketitle

\label{firstpage}

\begin{abstract}
The density and temperature structures of dense cores in the L1495 cloud of the 
Taurus star-forming region are investigated using {\it Herschel\/} SPIRE and 
PACS images in the 70 $\mu$m, 160 $\mu$m, 250 $\mu$m, 350 $\mu$m and 500 $\mu$m continuum bands. A sample consisting of 20 cores, selected using spectral and spatial
criteria, is analysed using a new maximum likelihood technique, COREFIT, which
takes full account of the instrumental point spread functions.
We obtain central dust temperatures, $T_0$, in the range 6--12 K
and find that, in the majority of cases, the radial density falloff at large 
radial distances is consistent with the asymptotic $r^{-2}$ variation expected 
for Bonnor-Ebert spheres. Two of our cores exhibit a significantly steeper
falloff, however, and since both appear to be gravitationally unstable,
such behaviour may have implications for collapse models.  We find a strong
negative correlation between $T_0$ and peak column density,
as expected if the dust is heated predominantly by the interstellar radiation
field.
At the temperatures we estimate for the core centres, carbon-bearing
molecules freeze out as ice mantles on dust grains, and this behaviour is 
supported here by
the lack of correspondence between our estimated core locations and the
previously-published positions of H$^{13}$CO$^+$ peaks.  On this basis,
our observations suggest a sublimation-zone radius typically $\sim10^4$ AU.
Comparison with previously-published N$_2$H$^+$ data at 8400 AU resolution, 
however, shows no evidence for N$_2$H$^+$ depletion at that resolution.
\end{abstract}

\begin{keywords}
stars: formation --- stars: protostars --- ISM: clouds --- submillimetre: ISM --- methods: data analysis --- techniques: high angular resolution.
\end{keywords}

\section{Introduction}
A key step in the star formation process is the production of cold dense
cores of molecular gas and dust \citep{wt94,andre1996}. 
Cores which do not contain a stellar object are referred to as
{\it starless\/}; an important subset of these 
consists of {\it prestellar\/} cores, i.e., those cores which are
gravitationally-bound and therefore present the initial conditions
for protostellar collapse. 

Observations of cold cores are best made in the submillimetre regime in
which they produce their peak emission, and observations made with
ground-based telescopes have previously helped to
establish important links between the stellar initial mass function (IMF)
 and the core mass function (CMF)
\citep{motte98}. With the advent of {\it Herschel\/} \citep{pilb10}, however,
these cores can now be probed with high-sensitivity multiband imaging
in the far infrared and submillimetre, and hence
the CMF can be probed to lower masses than before. One of the major goals
of the {\it Herschel\/} Gould Belt Survey \citep{andre10} is to characterise
the CMF over the densest portions of the Gould Belt. This survey covers
15 nearby molecular clouds which span a wide range of star formation
environments;  preliminary results for 
Aquila have been reported by \cite{kon10}. Another {\it Herschel\/}
key programme, HOBYS (``{\it Herschel\/} imaging survey of OB Young Stellar
objects") \citep{motte2010}, is aimed at more massive dense cores and the
initial conditions for high-mass star formation,
and preliminary results have been presented by \citet{gian12}.

The Taurus Molecular Cloud is a nearby region of predominantly non-clustered 
low-mass star formation, at an estimated distance of 140 pc \citep{elias78},
in which the stellar density is relatively low and objects can be studied
in relative isolation.  Its detailed morphology at {\it Herschel\/} wavelengths
is discussed by \cite{kirk2013}. The region is dominated by two long
($\sim 3^\circ$), roughly parallel filamentary structures, the larger
of which is the northern structure.  Early results from {\it Herschel\/}
regarding the filamentary properties have been reported by \cite{palm13}.  

In this paper we focus on the starless core population of the field with
particular interest in core structure and star-forming potential.
Our analysis is based on observations of the
western portion of the northern filamentary structure,
designated as N3 in \cite{kirk2013}, which includes the Lynds cloud L1495
and contains Barnard clouds B211 and B213 as prominent subsections of
the filament. Our analysis involves a sample of 20 cores which we believe to be 
representative of relatively isolated cores in this region. The principal
aims of the study are as follows: 
\begin{enumerate}
 \item  accurate mass estimation based on models which take account of
radial temperature variations and which use spatial and spectral data;
 \item  a comparison of these results with those from simpler techniques
commonly used for estimating the core mass function in order to provide
a calibration benchmark for such techniques;
 \item  investigation of processes such as heating of the dust by the 
interstellar radiation field (ISRF) and the effect 
of temperature gradients on core stability;
 \item examination of the results in the context of other observations of the
same cores where possible, particularly with regard to gaining insight into
the relationship between the dust and gas.
\end{enumerate}

The estimation of the core density and temperature structures is
achieved using our newly developed technique, COREFIT, 
complementary in some ways to the recently used Abel transform method 
\citep{roy2013}. Before discussing COREFIT and its results in detail, we first
describe our observations and core selection criteria.

\section[]{Observations}

The observational data for this study consists of a set of images of 
the L1495 cloud in the Taurus star-forming region, made on 12 February, 2010
and 7--8 August, 2010, during the course
of the {\it Herschel\/} Gould Belt Survey (HGBS).  The data were taken using
PACS \citep{pog10} at 70 $\mu$m and 160 $\mu$m and SPIRE \citep{griffin10} 
at 250 $\mu$m, 350 $\mu$m, and 500 $\mu$m in fast-scanning (60 arcsec/s) 
parallel mode.  The {\it Herschel\/} 
Observation IDs were 1342202254, 1342190616, and 1342202090.
An additional PACS observation (ID 1342242047) was taken
on 20 March 2012 to fill a data gap.
Calibrated scan-map images were produced in the HIPE Version 8.1 pipeline 
\citep{ott2010} using the Scanamorphos \citep{rous13}  and 
``naive" map-making procedures for PACS and SPIRE, 
respectively. A detailed description of the observational and data reduction 
procedures is given in \citet{kirk2013}. 

\section[]{Candidate core selection}

The first step in our core selection procedure consists
of source extraction via the {\it getsources\/} algorithm\footnote{Version 1.130401 was used for the analysis described here} \citep{men12} which
uses the images at all available wavelengths simultaneously.  These consist
of the images at all five {\it Herschel\/} 
bands plus a column density map which is used
as if it were a sixth band, the purpose being to
give extra weight to regions of high column density in the detection process.
The column density map itself is obtained from the same set of SPIRE/PACS
images, using the procedure described by \citet{palm13} which provides
a spatial resolution corresponding to that of the 250 $\mu$m observations.

The detection list is first filtered to remove unreliable sources. This
is based on the value of the ``global goodness" parameter \citep{men12} which 
is a combination of various quality metrics.  It incorporates the quadrature
sums of both the ``detection significance" and signal to noise ratio
($S/N$) over the set of wavebands, as well as some contrast-based information. 
The ``detection significance" is defined with respect to a spatially 
bandpass-filtered image, the characteristic spatial scale of which matches that
of the source itself.  At a given band, the detection significance is
then equal to the ratio of peak source intensity to the standard deviation
of background noise in this image. The $S/N$ is defined in a similar way,
except that it is based on the observed, rather than filtered,
image.

For present purposes we require a ``global goodness" value greater than or 
equal to 1.
A source satisfying this criterion may be regarded as having
an overall confidence level $\ge7\sigma$ and can therefore
be treated as a robust detection.
Classification as a core for the purpose of this study then involves the 
following additional criteria:

\begin{enumerate}
  \item detection significance (as defined above) greater than or equal to 5.0
 in the column density map;
  \item detection significance  greater than or equal to 5.0
in at least {\it two\/} wavebands between 160 $\mu$m
and 500 $\mu$m;
  \item detection significance {\it less than\/} 5.0 for the 70 $\mu$m band
and no visible signature on the 70 $\mu$m image, 
in order to exclude protostellar cores, i.e., those cores which contain
a protostellar object;
  \item ellipticity less than 2.0, as measured by {\it getsources\/};
  \item source not spatially coincident with a known galaxy, based on
comparison with the NASA Extragalactic Database.
\end{enumerate}

\begin{figure}
\includegraphics[width=84mm]{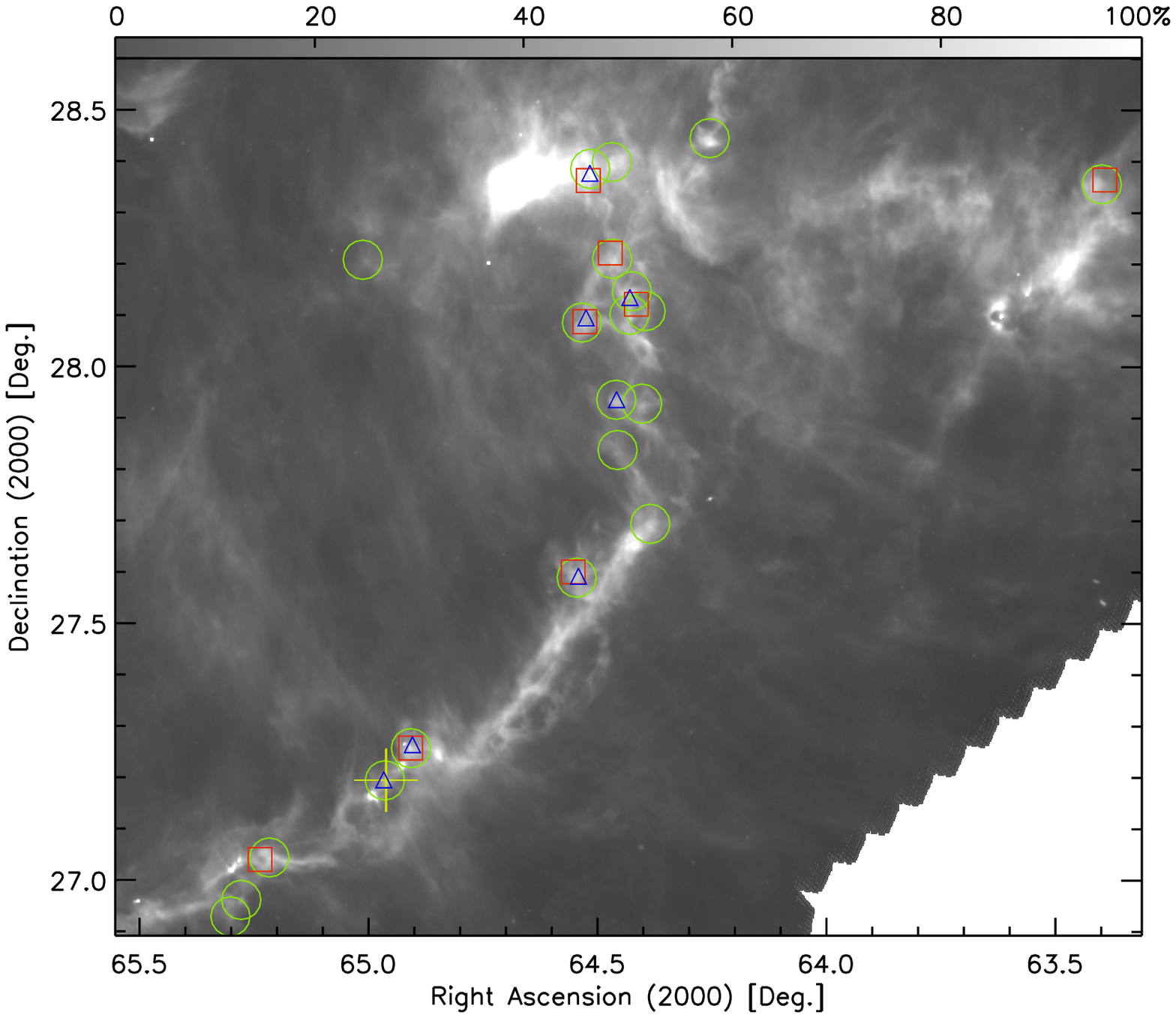}
 \caption{SPIRE 250 $\mu$m image of the L1495 region. The green circles
represent the locations of the 20 cores selected for modeling. The other
symbols represent previously published core locations at other wavelengths; 
{\it red squares\/}: H$^{13}$CO$^+$ \citep{onishi2002}; 
{\it blue triangles\/}: N$_2$H$^+$ \citep{hac2013}; {\it yellow cross\/}:
850 $\mu$m \citep{sad2010}. The image is shown on a truncated intensity scale
in order to emphasize faint structure; the display saturates at
200 MJy sr$^{-1}$ which corresponds to 100\% on the greyscale.}
 \label{fig1}
\end{figure}

This procedure resulted in a total of 496 cores over the observed 
$2^\circ\!\!.2 \times 2^\circ\!\!.2$ region. The total mass, 
88 $M_\odot$, of the detected cores represents approximately 4\% of the mass 
of the L1495 cloud, estimated to be 1500--2700 $M_\odot$ \citep{kram91}. 
From this set, 20 cores were selected for detailed study.
The main goal of the final selection process was to obtain a list of relatively
unconfused cores, uniformly sampled in mass according to preliminary 
estimates obtained via SED fitting as outlined
in the next section. Cores which were multiply peaked or confused, based on
visual examination of the 250--500 $\mu$m images, were excluded. The mass 
range 0.02--2.0 $M_\odot$ was then
divided into seven bins, each of which spanned a 
factor of two in mass, and a small number of objects (nominally three) 
selected from each bin. The selection was made on a random basis except for 
a preference for objects for which previously-published data were available, 
thus facilitating comparison of deduced parameters. 
Fig. \ref{fig1} shows the locations of the 20 selected cores on a 
SPIRE 250 $\mu$m image of the field. 

\section[]{SED fitting}

Preliminary values of core masses and dust temperatures are estimated 
by fitting a greybody spectrum to the observed spectral
energy distribution (SED) constructed from the set of five-wavelength 
{\it getsources\/} fluxes.  For this computation, sources are assumed to be 
isothermal and have a wavelength variation of opacity of the form 
\citep{hild83,roy2013}:
\begin{equation}
    \kappa(\lambda) = 0.1\,(300/\lambda_{[\mu{\rm m}]})^2\qquad{\rm cm}^2\,{\rm g}^{-1}
    \label{eq1}
\end{equation}
Although obtained observationally, the numerical value of the coefficient in 
this relation is consistent with a gas to dust ratio of 100.

\section[]{Core profiling}

To obtain better estimates of core mass and other properties, a
more detailed model fit is required.
For this purpose we have developed a new procedure, COREFIT, 
which involves maximum
likelihood estimation using both spatial and spectral information.

The fitting process involves calculating a series of forward models,
i.e., sets of model images based on assumed parameter values, which are then
compared with the data.  The models
are based on spherical geometry,
in which the radial variations of volume density and temperature are
represented by parametrized functional forms. For a given set of parameters,
a model image is generated at each of the five wavelengths by calculating
the emergent intensity distribution on the plane of the sky and convolving
it with the instrumental point spread function (PSF)\footnote{For the PACS 
images, we use azimuthally-averaged versions of the PSFs estimated from
observations of Vesta \citep{lutz2012}; for SPIRE we use 
rotationally symmetric PSFs based on the measured radial profiles presented by 
\citet{griffin13}.} at the particular wavelength.
The parameters are then adjusted to obtain an inverse-variance weighted
least squares fit to the observed images.

In this procedure the wavelength variation of opacity is assumed to be given
by Eq. (\ref{eq1}) and the 
the radial variations of volume density and dust temperature are assumed
to be described by Plummer-like \citep{plum11} and quadratic forms, 
respectively. Specifically we use:
\begin{eqnarray}
    n(r) & = & n_0/[1 + (r/r_0)^\alpha]  \label{eq2} \\
    T(r) & = & T_0 + (T_1\!-\!T_0\!-\!T_2)r/r_{\rm out} + T_2(r/r_{\rm out})^2
    \label{eq3}
\end{eqnarray}
where $n(r)$ represents the number density of H$_2$ molecules at radial
distance $r$, $r_0$ represents the radius of an inner
plateau, and $r_{\rm out}$ is the outer radius of the core, outside of which
the core density is assumed to be zero. Also, $T_0$ is the central core
temperature, $T_1$ is the temperature at the outer radius,
and $T_2$ is a coefficient which determines 
the curvature of the radial temperature profile. 
In relating $n(r)$ to the corresponding profile of mass 
density we assume a mean molecular weight of 2.8 \citep{roy2013}. 

The set of unknowns then consists of:
$n_0$, $r_0$, $r_{\rm out}$, $\alpha$, $T_0$, $T_1$, $T_2$,
$x$, $y$, where the latter two variables represent the angular coordinates
of the core centre.  Representing this set by
an 9-component parameter vector, {\bf p}, we can write the measurement model
as:
\begin{equation}
    {\bf \zeta}_\lambda = {\bf f}_\lambda({\bf p}) + b_\lambda + \nu_\lambda
    \label{eq4}
\end{equation}
where $\zeta_\lambda$ is a vector representing the set of pixels of the 
observed image at wavelength $\lambda$, ${\bf f}_\lambda({\bf p})$ represents
the model core image for parameter set {\bf p}, and $\nu_\lambda$ is the 
measurement noise vector, assumed to be an uncorrelated zero-mean Gaussian 
random process.  
Also, $b_\lambda$ represents the local background level, estimated using
the histogram of pixel values in an annulus\footnote{The inner radius of this
annulus is taken as the size of the source ``footprint" which is estimated by  
{\it getsources\/} and includes all of the source emission on the observed
images; the outer radius is set 10\% larger.} surrounding the source.  This 
measurement model assumes implicitly that the core is optically thin
at all wavelengths of observation.

In principle, the solution procedure is then to minimise the chi squared
function, $\phi({\mathbf p})$, given by:
\begin{equation}
    \phi({\mathbf p}) = \sum_{\lambda,i} [(\zeta_\lambda)_i
-b_\lambda - ({\bf f}_\lambda({\bf p}))_i]^2 / \sigma_\lambda^2 
    \label{eq5}
\end{equation}
where subscript $i$ refers to the $i$th pixel of the image at the given
wavelength and
$\sigma_\lambda$ represents the standard deviation of the measurement errors,
evaluated from the sky background fluctuations in the background annulus.
\smallskip

In practice, two difficulties arise:

\begin{enumerate}

\item An unconstrained minimisation of
$\phi({\mathbf p})$ is numerically unstable due to the fact that for a
given total number of molecules, $n_0$ in Eq. (\ref{eq2}) becomes infinite as
$r_0\rightarrow 0$. It results in
near-degeneracy such that the data do not serve to distinguish between a
large range of possible values of the central density. To overcome this, 
we have 
modified the procedure to incorporate the constraint $r_0\ge r_{\rm min}$,
where $r_{\rm min}$ is equal to one quarter of
the nominal angular resolution, which we take to be the beamwidth at
250 $\mu$m.  The estimate of central density then becomes a
``beam-averaged" value over a resolution element of area
$\pi r_{\rm min}^2$.  For a distance of 140 pc, $r_{\rm min}$ 
corresponds to about 600 AU.  

\item Most cores show some degree of asymmetry. This can degrade the
quality of the global fit to a spherically-symmetric model, causing the
centre of symmetry to miss the physical centre of the core. Some negative
consequences include
an underestimate of the central density and an overestimate of the central
temperature. To alleviate this, we estimate the $(x,y)$ location of the core
centre ahead of time using the peak of a column density map, constructed
at the spatial resolution of the 250 $\mu$m image. The maximum
likelihood estimation is then carried out using a 7-component parameter vector
which no longer involves the positional variables.

\end{enumerate}

Having performed the position estimation and constrained chi squared
minimisation,
the core mass is then obtained by integrating the density profile given
by Eq. (\ref{eq2}), evaluated using the estimated values of $n_0$, $r_0$
and $\alpha$.

Evaluation of the uncertainties in parameter estimates is complicated by
the nonlinear nature of the problem which leads to a multiple-valley nature 
of $\phi({\mathbf p})$. The usual procedure, in which the uncertainty
is evaluated by inverting a matrix of 2nd derivatives of 
$\phi({\mathbf p})$ \citep{wh71}, then only provides values which correspond
to the width of the global maximum and ignores the existence of
neighbouring peaks which may represent significant probabilities.
We therefore evaluate the uncertainties using a Monte Carlo technique
in which we repeat the estimation procedure after adding a series of
samples of random noise to the observational data and examine the effect on the
estimated parameters.

We have also implemented an alternate version of COREFIT, referred to
as ``COREFIT-PH," in which the dust 
temperature profile is based on a radiative transfer model, PHAETHON 
\citep{stam03}, rather than estimating it from the observations.  
In this model, the radial density profile has the same form as for
the standard COREFIT (Eq. (\ref{eq2})) but with the index, $\alpha$,
fixed at 2. 
The temperature profile is assumed to be determined entirely by the 
heating of dust by the external ISRF; the latter is modeled locally as
a scaled version of the standard ISRF \citep{stam07} using a 
scaling factor, $\chi_{\rm ISRF}$, which represents an additional variable
in the maximum likelihood solution.

We now compare the results 
obtained using the two approaches, both for synthetic and real data.

\subsection[]{Tests with synthetic data}

We have tested both COREFIT and COREFIT-PH against synthetic data 
generated using an alternate forward model for dust radiative transfer, 
namely MODUST (Bouwman et al., in preparation).  Using the latter code,
images at the five wavelengths were generated for a
set of model cores and convolved with Gaussian simulated PSFs with full width 
at half maxima (FWHM) corresponding to the {\it Herschel\/} beamsizes.  
The models involved central number densities of $10^5$ cm$^{-3}$, 
$10^6$ cm$^{-3}$, and $3\times10^6$ cm$^{-3}$ with corresponding $r_0$ values 
of 2500 AU, 4000 AU, and 1000 AU, respectively, and $r_{\rm out}$ values of
$1.3\times10^4$ AU, $1.7\times10^4$ AU, and $1.2\times10^4$ AU, respectively. 
The corresponding core masses
were 0.59 $M_\odot$, 18.37 $M_\odot$, and 3.11 $M_\odot$, respectively.
The synthesized 
images and corresponding Gaussian PSFs were then used as input data to
the inversion algorithms.  The results are presented in Table 1.

It is apparent that COREFIT gave masses and
central temperatures in good agreement with the true model.  While 
COREFIT-PH reproduced the central temperatures equally well, 
it underestimated the masses of these simulated cores by factors of 
0.7, 0.5, and 0.5, respectively. The reason for these differences is that 
even though the two radiative transfer codes (PHAETHON and MODUST)
yield central temperatures in good agreement with each other for a given
set of model parameters, they produce divergent results for the
dust temperatures in the outer parts of the cores, due largely to differences
in dust model opacities.  Since the
outer parts comprise a greater fraction of the mass than does the central
plateau region, this can lead to substantially different mass estimates given
the same data.  This problem does not occur for COREFIT
since the latter obtains the temperature largely from the spectral
variation of the data rather than from a physical model
involving additional assumptions. These calculations 
thus serve to illustrate the advantages of simultaneous estimation of the
radial profiles of dust temperature and density.

\subsection[]{Results obtained with observational data}

\begin{figure}
\includegraphics[width=88mm]{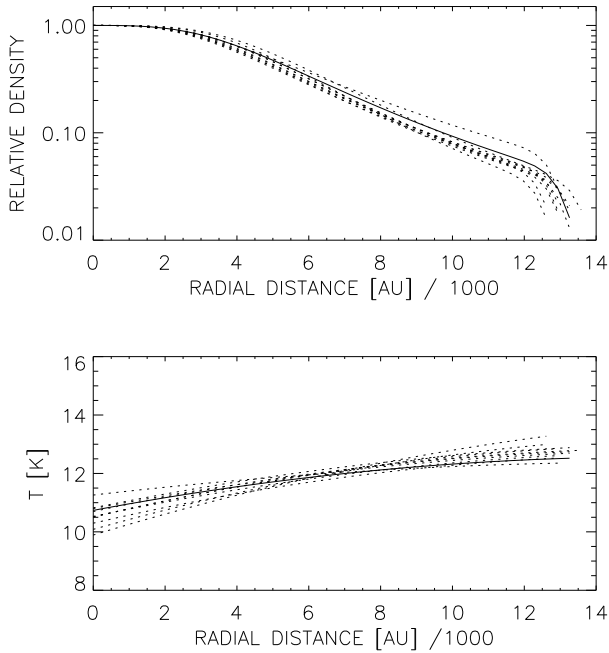}
 \caption{Example of COREFIT results, for a 0.7 $M_\odot$ core in 
L1495 (No.~2 in Table 2).
The solid lines indicate maximum likelihood estimates of
the profiles of relative volume density and dust temperature.
The dashed lines provide a measure of the
uncertainty in the estimated density and temperature. They represent the 
results of a Monte Carlo simulation in which the estimation procedure is
repeated 10 times after adding synthetic measurement noise to the
observed images; the standard deviation of the added noise corresponds to
the estimated measurement noise of the observed images.}
 \label{fig2}
\end{figure}

\begin{figure}
\includegraphics[width=88mm]{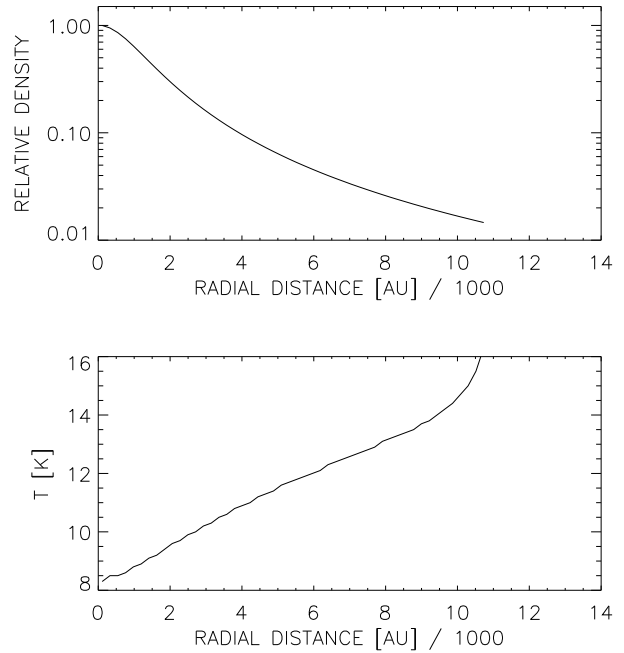}
 \caption{COREFIT-PH results for the same core as in Fig. \ref{fig2}.
In this variant of the estimation procedure,
the dust temperature profile is modeled using a radiative transfer code
(PHAETHON) instead of estimating it from the observations. }
 \label{fig3}
\end{figure}

Table 2 shows the complete set of COREFIT parameter estimates for each of the 
Taurus cores. Also included are the assumed values of the inner radius
of the annulus used for background estimation, equal to the {\it getsources\/} 
footprint size.  Table 3 shows a comparison of the
mass and temperature estimates amongst the different techniques,
which include COREFIT and COREFIT-PH as well as the SED fitting 
discussed in Section 3. To facilitate comparison between the COREFIT
temperatures and the mean core temperatures estimated from the
spatially integrated fluxes used in the SED fits, 
we include the spatially averaged COREFIT temperature,
$\bar{T}$, defined as the density-weighted mean value of 
$T(r)$ for $r\le r_{\rm out}$.  The COREFIT-PH results include the 
values of the ISRF scaling factor, $\chi_{\rm ISRF}$, the median value
of which is 0.33.  The fact that this is noticeably less than unity
can probably be attributed to the fact that these cores
are all embedded in filaments and hence the local ISRF is attenuated by
overlying filamentary material.
As an example of the fitting results, Figs. \ref{fig2} and \ref{fig3} show 
the estimated density and temperature profiles for core No.~2 in Table 2,
based on COREFIT and COREFIT-PH, respectively.

\begin{figure}
\includegraphics[width=88mm]{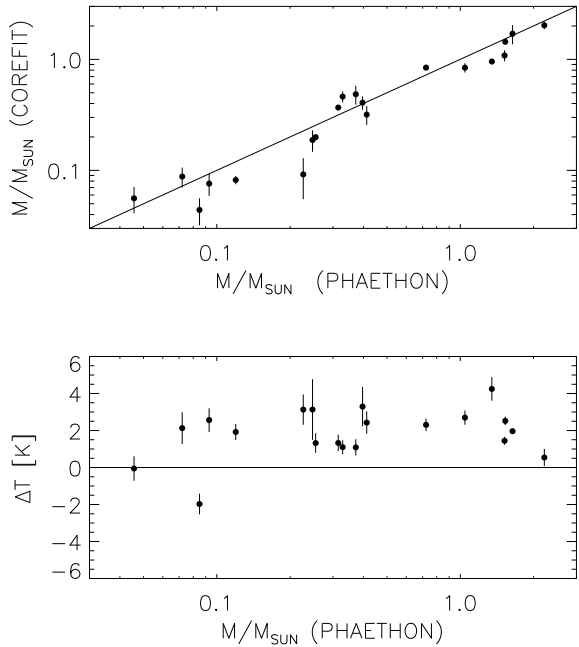}
 \caption{Comparison between parameter estimates, obtained using COREFIT
and COREFIT-PH, for the 20 selected cores. In the former 
procedure, the dust temperature profile is estimated directly from the 
observations, while in the latter it is modeled using radiative transfer.
{\it Upper plot\/}: COREFIT mass versus COREFIT-PH mass. For reference,
the solid line represents the locus of equal masses.
{\it Lower plot\/}:  $\Delta T$ versus mass, where
$\Delta T$ represents the difference in estimated temperature 
(COREFIT minus COREFIT-PH).}
 \label{fig4}
\end{figure}

Fig. \ref{fig4} shows that the two techniques yield consistent
estimates of masses, but the radiative transfer calculations produce
central temperatures which are, on average, $\sim2$ K lower than the COREFIT 
estimates. Although the difference is not significant in individual cases
(the standard deviation being 1.4), it is clear from Fig. \ref{fig4} that a 
systematic offset is present; the mean temperature difference, $\Delta T$, is
$1.9\pm0.3$ K.

Based on the results of testing with synthetic data, this difference
seems too large to be explained by systematic errors associated with dust grain
models, although we cannot rule out that possibility. One could also question
whether our $\chi_{\rm ISRF}$ values are spuriously low. We do, in fact,
find that by forcing the latter parameter to a somewhat larger value (0.5), 
the median 
$\Delta T$ can be reduced to zero with only a modest increase in the reduced
chi squared, $\chi_\nu^2$ (0.85 as opposed to 0.83 for the best fit). The
observations are completely inconsistent with $\chi_{\rm ISRF}=1.0$, however.
As an additional test, we can take the COREFIT estimate of the radial 
density distribution for each core and
use the standalone PHAETHON code to predict the central temperature for any
assumed value of $\chi_{\rm ISRF}$.  We thereby obtain consistency with the 
COREFIT estimates with $\chi_{\rm ISRF}=0.67$.  However, this consistency
comes at significant cost in terms of goodness of fit (the median 
$\chi_\nu^2$ increases to 2.27), and therefore does {\it not\/} serve 
to reconcile the COREFIT results with the expectations of radiative transfer. 
In summary, the COREFIT results are not entirely consistent with our assumed 
model for dust heating by the ISRF, but further work will be necessary to 
determine whether the differences are model-related or have astrophysical 
implications.  So at this stage we have no 
evidence to contradict the findings of \citet{evans01} who considered 
various heating sources (the primary and secondary effects of cosmic rays
and heating of dust grains by collisions with warmer gas particles) and 
concluded that heating by the ISRF dominates over all other effects.

\begin{figure}
\includegraphics[width=88mm]{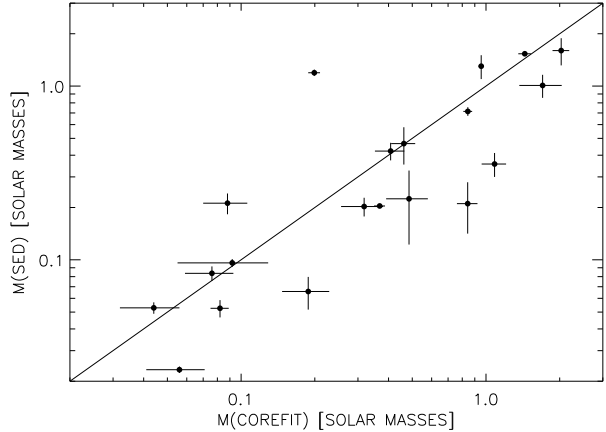}
 \caption{SED-derived mass based on isothermal assumption versus the mass from
COREFIT model. For reference, the solid line represents the locus of equal
masses.}
 \label{fig5}
\end{figure}

How do the COREFIT estimates of temperature and mass compare 
with the preliminary values estimated from the {\it getsources\/} SEDs?  
In the case of temperature, the relevant comparison is between the 
SED-derived value and the spatially averaged COREFIT value;
the data in Table 3 then give a mean ``COREFIT minus SED" difference of
-0.2 K, with a standard deviation of 1.1 for individual cores. The temperature
estimates are thus consistent.  With regard to mass,
Fig. \ref{fig5} shows that SED
fitting under the isothermal assumption yields masses that are systematically
smaller than the COREFIT values; 
the mean ratio of COREFIT mass to SED-based mass is 1.5, with
a standard deviation of 1.0 in individual cases.
Since the internal temperature
gradient increases with the core mass, one might expect that the
correction factor for SED-derived masses would increase with mass,
although Fig. \ref{fig5} has too much scatter to establish this.
It may be evident
when the results are averaged for a much larger statistical sample of cores,
although the correction may well depend on environmental factors such as
the intensity of the local ISRF.

\begin{figure}
\includegraphics[width=88mm]{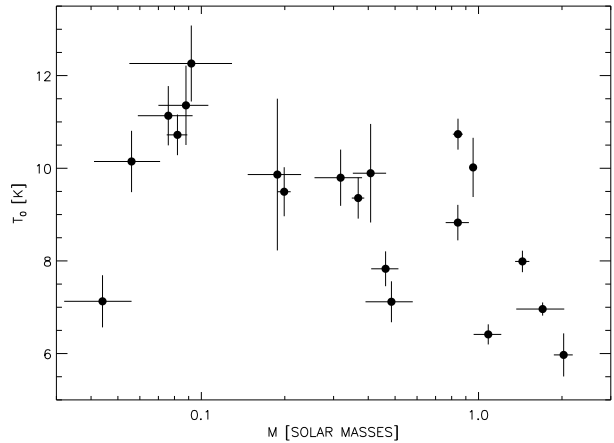}
 \caption{Central dust temperature, $T_0$, as a function of core mass.}
 \label{fig6}
\end{figure}

\begin{figure}
\includegraphics[width=88mm]{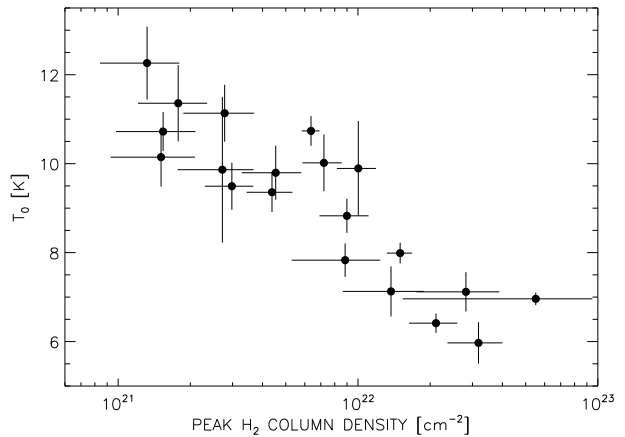}
 \caption{Central dust temperature, $T_0$, as a function of peak column 
density of H$_2$ molecules.}
 \label{fig7}
\end{figure}

Fig. \ref{fig6} shows a plot of estimated central temperature as
a function of core mass. Linear regression indicates that these quantities
are negatively correlated with a coefficient of -0.64. This correlation
can be explained quite naturally as a consequence of increased shielding of 
the core, from the ISRF, with increasing core mass. This being the case,
one would expect an even stronger correlation
with peak column density and this is 
confirmed by Fig. \ref{fig7}, for which the associated 
correlation coefficient is -0.86.

\begin{figure}
\includegraphics[width=88mm]{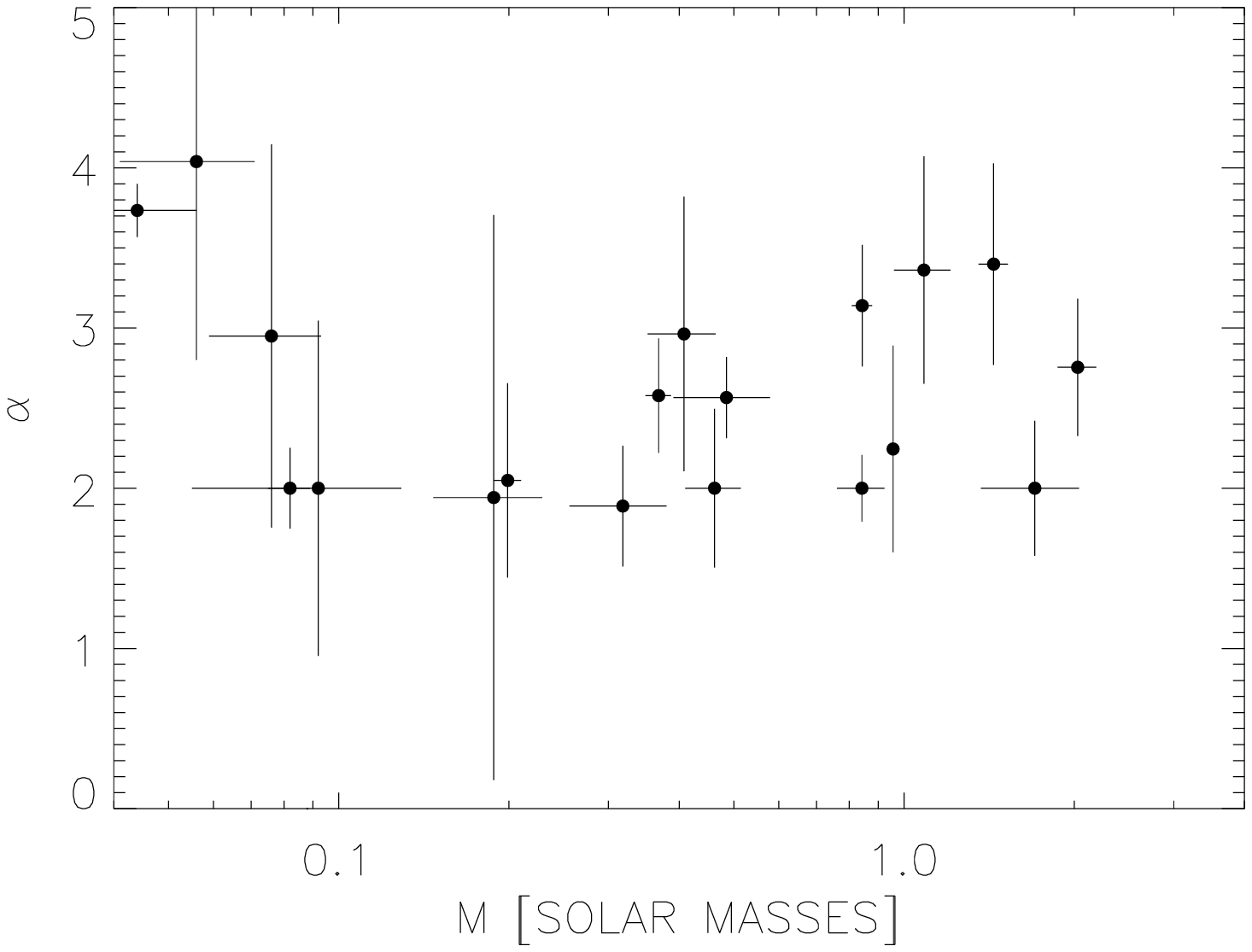}
 \caption{The estimated index of radial density falloff, $\alpha$,
as a function of core mass.}
 \label{fig8}
\end{figure}

Fig. \ref{fig8} shows a plot of $\alpha$ versus mass, 
where $\alpha$ is the index of radial
density variation as defined by Eq. (\ref{eq2}), and the masses are the COREFIT
values.  Given the relatively large uncertainties, the $\alpha$ values are,
for the most part, consistent with values expected for Bonnor-Ebert spheres,
whereby $\alpha=2.5$ provides an accurate empirical 
representation at radial distances up to the instability radius \citep{taf04},
and that $\alpha$ decreases to its asymptotic value of 2 beyond that.

\begin{figure}
\hspace*{-0.7cm}\includegraphics[width=95mm]{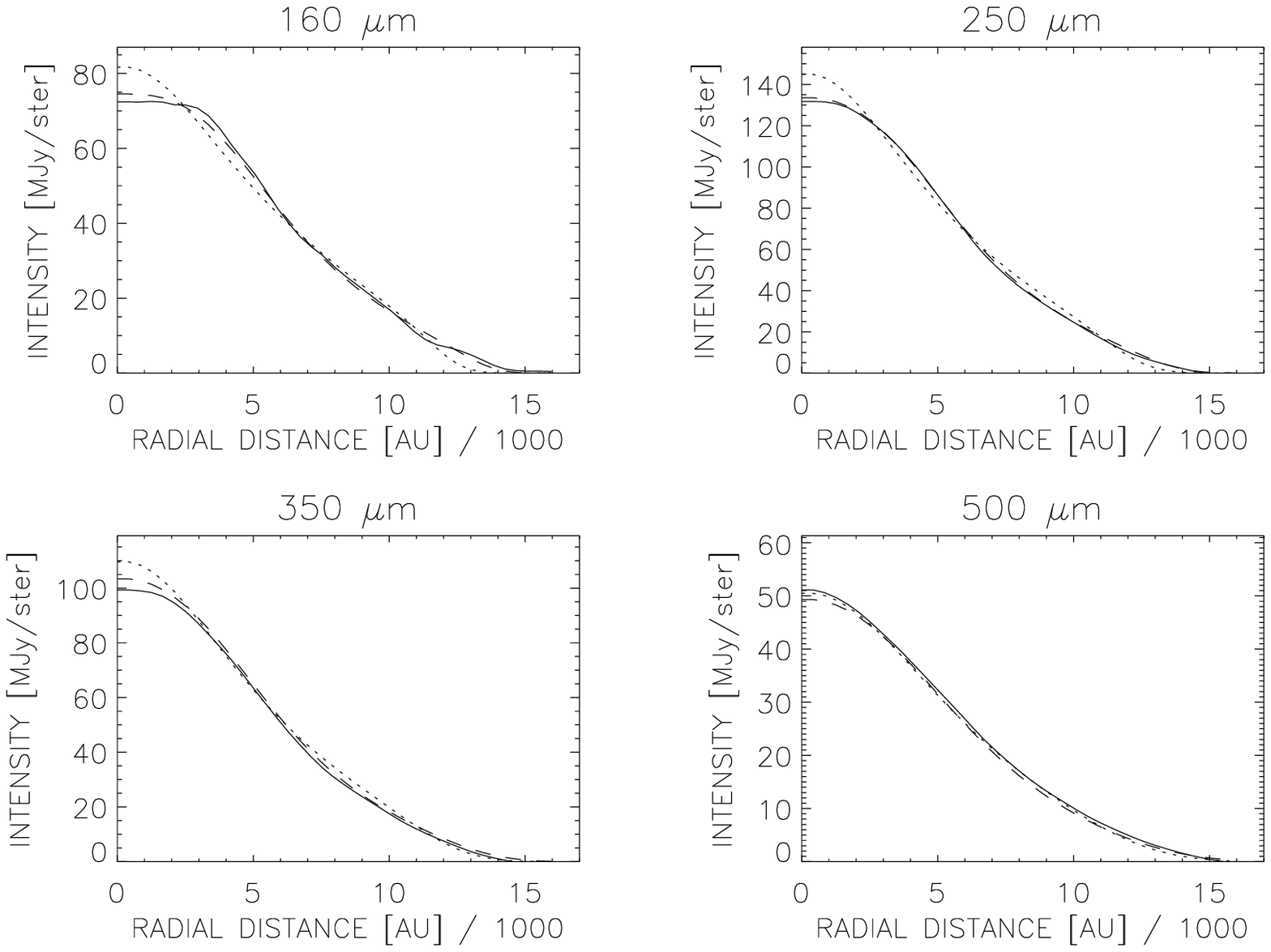}
 \caption{Radial profiles of the images of core No. 2 at four wavelengths,
showing the match between observations and models for two different values of 
the radial density index, $\alpha$.  {\it Solid line:\/} observed
profile;  {\it dashed line:\/} best fitting model ($\alpha=3.1$) convolved 
with the corresponding PSF at each wavelength);
{\it dotted line:\/} same, except for the constraint $\alpha\!=\!2.0$. Note that
the latter model results in a poor fit in the central region.}
 \label{fig9}
\end{figure}

The general consistency with the Bonnor-Ebert model is supported
by the fact that when the maximum likelihood fitting procedure is repeated
using the constraint $\alpha=2$, the chi squared values are, in most cases,
not significantly different from the values obtained when $\alpha$ is
allowed to vary.  Two exceptions, however, are cores 2 and 13, both of which
are fit significantly better by density profiles steeper than Bonnor-Ebert
($\alpha=3.1$ and 2.8, respectively), as illustrated by Fig. \ref{fig9} for
the former case.  Specifically, the 
chi squared\footnote{To evaluate this quantity, the number of degrees
of freedom, $N_{\rm df}$, was taken as the total number of resolution elements
contained within the fitted region for all five input images; 
$N_{\rm df}$ is then $\sim1700$ and $\sim1200$ for the two cases, 
respectively.} differences 
(17.2 and 7.5, respectively) translate into relative probabilities, for 
the ``$\alpha\!=\!2$" hypothesis, of $\sim 2\times 10^{-4}$ and 0.02, 
respectively.  If confirmed, such behaviour may
have some important implications for core collapse models; a steepening of
the density distribution in the early collapse phase is, in fact, predicted
by the model of \citet{voro2005} in which the collapsing core begins to
detach from its outer boundary.  

\section[]{Core stability}

Assessments of core stability are frequently made using SED-based estimates
of core mass and temperature and observed source size, assuming that
cores are isothermal and can be described as Bonnor-Ebert spheres
\citep{lada08}. Using the SED-based data in Table 3 in conjunction with the
{\it getsources\/} estimates of core size, we thereby find that the
estimated core mass exceeds the Bonnor-Ebert critical mass for 10 of
our 20 cores, suggesting that half of our cores are unstable to
gravitational collapse.

\begin{figure}
\includegraphics[width=88mm]{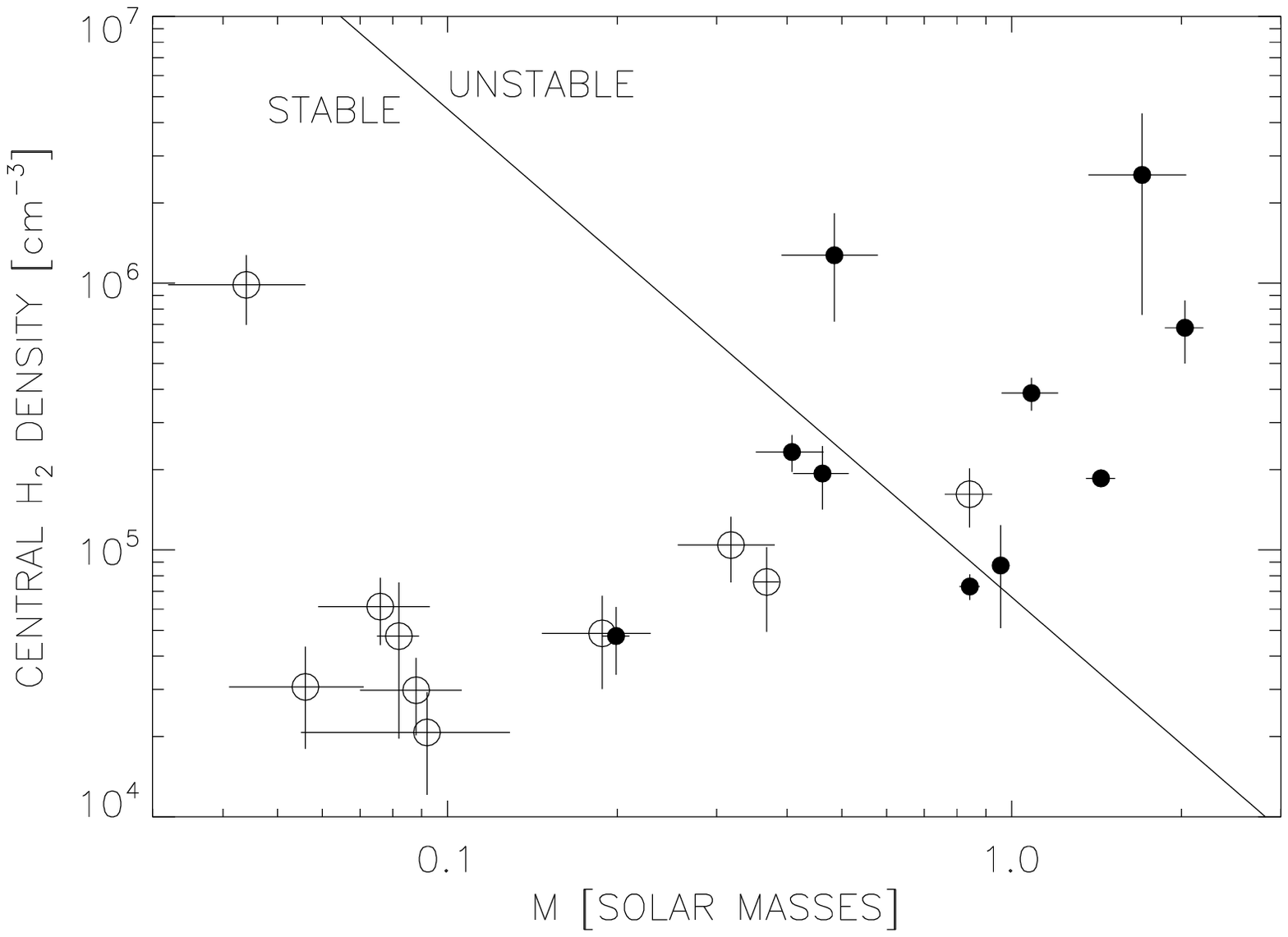}
 \caption{Central number density of H$_2$ molecules as a function of core mass.
The circles represent the COREFIT estimates for L1495; filled symbols
designate the subset of cores whose preliminary assessment of dynamical status
suggests that they are gravitationally bound, based on {\it getsources\/} 
fluxes and sizes in conjunction with the standard model of isothermal 
Bonnor-Ebert spheres.
For comparison, the solid line represents the locus of critically-stable
non-isothermal Bonnor-Ebert spheres (Sipil\"a, Harju \& Juvela (2011); points
to the right of this line represent cores which are unstable to
gravitational collapse according to that model.}
 \label{fig10}
\end{figure}

Our COREFIT parameter estimates enable us to make a more detailed assessment of
core stability based on a comparison with the results of hydrostatic
model calculations that take account of the
non-isothermal nature of the cores.  This is facilitated by the modified 
Bonnor-Ebert (MBE) sphere models of \citet{sip11}.
Adopting their model curves, based on the \cite{li01}
grains which best reproduce our estimated core temperatures,
the locus of critical non-isothermal models on a density versus mass plot
is shown by the solid line on Fig. \ref{fig10}.
Also plotted on this figure, for comparison, are the COREFIT estimates of 
those quantities. The seven points to the right of this curve
represent cores that we would consider to be gravitationally unstable
based on the modified Bonnor-Ebert models. Although this is somewhat less
than the 10 that were classified as unstable based on the SED fits,
the difference is probably not significant given that several points
on the plot lie close to the ``stability" line.

The consistency between the above two procedures for stability assessment
is illustrated by the fact that the MBE stability line in Fig. \ref{fig10}
provides a fairly clean demarcation between the 
cores classified as stable (open circles) and unstable (filled circles)
from the simpler (SED-based) procedure. These results therefore suggest 
that prestellar cores can be identified reliably as such using relatively 
simple criteria.

The Bonnor-Ebert model also provides a stability criterion with respect
to the centre-to-edge density contrast, whereby values greater than 14
indicate instability to gravitational collapse,
both for the isothermal and non-isothermal cases \citep{sip11}. However
the outer boundaries are not well defined for the present sample of cores, 
and consequently
the contrast values are uncertain in most cases. Two exceptions are cores 2 
and 13, both of which have contrast estimates whose significance exceeds 
$3\sigma$.  In both cases the mass exceeds the Bonnor-Ebert critical 
mass (by ratios of 1.2 and 6.0, respectively), and the 
centre-to-edge contrast values ($20\pm6$ and $104\pm32$, respectively) are in 
excess of 14. So for those two cores, at least, the core stability deduced from 
the density constrast is thus consistent with that assessed from total mass.

\begin{table*}
 \centering
 \begin{minipage}{140mm}
  \caption{Results of testing with synthetic data.}
  \begin{tabular}{@{}rccccccccccccccccc@{}}
  \hline
   & & &  \multicolumn{3}{c}{$T_0$ [K]} 
& & & \multicolumn{3}{c}{Mass [$M_\odot$]} 
& & & \multicolumn{3}{c}{Peak H$_2$ col. dens. [10$^{22}$ cm$^{-2}$]}\\
   Model & & & Std.\footnote{Standard version of COREFIT} & Alt.\footnote{Alternate version (COREFIT-PH)} & True & & & Std. & Alt. & True & & & Std.  & Alt. & True  \\
 \hline
  1 & & & 10.1 & 8.6 & 10.0 & & & 0.61 & 0.39 & 0.59 & & & 0.59 & 0.45 & 1.04 \\
  2 & & & 6.8 & 6.5 &  6.5 & & & 22.7 & 9.48& 18.4 & & &23.9 & 5.65 &16.1 \\
  3 & & & 7.8 & 6.5 &  6.7 & & & 3.30 & 1.46 & 3.11 & & & 7.95 & 4.91 & 13.4 \\
\hline
\end{tabular}
\end{minipage}
\end{table*}

\begin{table*}
 \centering
 \begin{minipage}{200mm}
  \caption{COREFIT parameter estimates for the 20 Taurus cores.}
  \begin{tabular}{@{}rccrccccccc@{}}
  \hline
   Core  & RA & Dec & $r_{\rm annulus}$\footnote{Inner radius of the annulus used for background estimation.} & $r_0$ & $r_{\rm out}$ & $n_0$ &$\alpha$
\footnote{An entry of `indet.' indicates that $\alpha$ is indeterminate from 
the data; this occurs if $r_0\simeq r_{\rm out}$.}
 & $T_0$\footnote{Central core temperature, effectively an average over a
resolution element of radius $r_{\rm min}=600$ AU.} 
 & $T_1$        & $T_2$\footnote{Negative values of $T_2$
indicate negative curvature of the temperature profile and have no
correspondence with actual temperatures.} \\
No. &(J2000)&(J2000) & [$10^3$ AU] &  [$10^3$ AU] & [$10^3$ AU]& [$10^5$ cm$^{-3}$] & & [K] & [K] & [K] \\
 \hline
  1 & 04:13:35.8 & +28:21:11 &   7.56 & $  3.5\pm  1.2$ & $  7.1\pm  0.2$ & $  0.5\pm  0.1$ & $  2.0\pm  0.6$ & $  9.5\pm  0.5$ & $ 11.2\pm  0.1$ & $ -1.4\pm  0.6$ \\
  2 & 04:17:00.6 & +28:26:32 &  16.38 & $  4.7\pm  0.2$ & $ 13.2\pm  0.6$ & $  0.7\pm  0.1$ & $  3.1\pm  0.4$ & $ 10.7\pm  0.3$ & $ 12.5\pm  0.2$ & $ -1.2\pm  0.2$ \\
  3 & 04:17:32.3 & +27:41:27 &  11.20 & $  3.5\pm  3.1$ & $  7.3\pm  0.4$ & $  0.2\pm  0.1$ & $  2.0\pm  1.0$ & $ 12.3\pm  0.8$ & $ 13.1\pm  1.2$ & $  0.2\pm  0.1$ \\
  4 & 04:17:35.2 & +28:06:36 &  10.30 & $  0.6\pm  0.2$ & $  8.4\pm  0.4$ & $  9.9\pm  2.9$ & $  3.7\pm  0.2$ & $  7.1\pm  0.6$ & $ 23.7\pm  1.2$ & $ -2.6\pm  0.5$ \\
  5 & 04:17:36.2 & +27:55:46 &  13.20 & $  2.9\pm  1.0$ & $ 11.8\pm  0.4$ & $  0.8\pm  0.3$ & $  2.6\pm  0.4$ & $  9.4\pm  0.4$ & $ 12.0\pm  0.3$ & $ -2.3\pm  0.5$ \\
  6 & 04:17:41.8 & +28:08:47 &  10.08 & $  0.8\pm  0.2$ & $  8.8\pm  1.2$ & $ 26\pm 18$ & $  2.0\pm  0.4$ & $  7.0\pm  0.1$ & $  8.3\pm  0.4$ & $ -1.1\pm  0.3$ \\
  7 & 04:17:43.2 & +28:05:59 &   7.14 & $  1.0\pm  0.4$ & $  6.9\pm  0.2$ & $ 13\pm 6$ & $  2.6\pm  0.3$ & $  7.1\pm  0.4$ & $ 10.1\pm  0.2$ & $ -2.9\pm  0.5$ \\
  8 & 04:17:49.4 & +27:50:13 &   8.10 & $  2.9\pm  2.9$ & $  7.7\pm  0.2$ & $  0.5\pm  0.2$ & $  1.9\pm  1.8$ & $  9.9\pm  1.6$ & $ 10.8\pm  0.5$ & $  0.9\pm  0.6$ \\
  9 & 04:17:50.6 & +27:56:01 &  13.86 & $  4.5\pm  0.2$ & $ 10.4\pm  0.4$ & $  1.9\pm  0.1$ & $  3.4\pm  0.6$ & $  8.0\pm  0.2$ & $ 10.1\pm  0.2$ & $ -1.0\pm  0.2$ \\
 10 & 04:17:52.0 & +28:12:26 &  12.60 & $  4.3\pm  1.7$ & $ 11.2\pm  0.2$ & $  0.9\pm  0.4$ & $  2.2\pm  0.6$ & $ 10.0\pm  0.6$ & $  9.7\pm  0.6$ & $ -0.1\pm  0.3$ \\
 11 & 04:17:52.5 & +28:23:43 &  12.15 & $  2.1\pm  1.0$ & $  9.5\pm  0.2$ & $  1.0\pm  0.3$ & $  1.9\pm  0.4$ & $  9.8\pm  0.6$ & $ 11.2\pm  0.3$ & $ -0.4\pm  0.2$ \\
 12 & 04:18:03.8 & +28:23:06 &   8.75 & $  2.3\pm  0.6$ & $  8.6\pm  0.2$ & $  2.3\pm  0.4$ & $  3.0\pm  0.9$ & $  9.9\pm  1.1$ & $ 10.8\pm  1.1$ & $  0.9\pm  0.7$ \\
 13 & 04:18:08.4 & +28:05:12 &  13.86 & $  2.3\pm  0.8$ & $ 13.4\pm  0.2$ & $  6.8\pm  1.8$ & $  2.8\pm  0.4$ & $  6.0\pm  0.5$ & $ 11.4\pm  0.3$ & $ -3.2\pm  0.5$ \\
 14 & 04:18:11.5 & +27:35:15 &   7.45 & $  2.3\pm  1.4$ & $  7.3\pm  0.2$ & $  1.9\pm  0.5$ & $  2.0\pm  0.5$ & $  7.8\pm  0.4$ & $ 10.1\pm  0.3$ & $ -2.2\pm  0.7$ \\
 15 & 04:19:37.6 & +27:15:31 &  11.76 & $  2.7\pm  0.6$ & $ 10.7\pm  0.4$ & $  1.6\pm  0.4$ & $  2.0\pm  0.2$ & $  8.8\pm  0.4$ & $ 10.6\pm  0.1$ & $ -1.5\pm  0.2$ \\
 16 & 04:19:51.7 & +27:11:33 &   9.24 & $  3.1\pm  0.4$ & $  8.9\pm  0.2$ & $  3.9\pm  0.5$ & $  3.4\pm  0.7$ & $  6.4\pm  0.2$ & $  9.6\pm  0.3$ & $ -3.0\pm  0.5$ \\
 17 & 04:20:02.9 & +28:12:26 &  10.30 & $  1.4\pm  0.4$ & $ 10.0\pm  0.4$ & $  0.5\pm  0.3$ & $  2.0\pm  0.3$ & $ 10.7\pm  0.4$ & $ 12.7\pm  0.2$ & $ -1.9\pm  0.5$ \\
 18 & 04:20:52.5 & +27:02:20 &   5.30 & $  5.2\pm  2.9$ & $  5.2\pm  0.2$ & $  0.3\pm  0.1$ &    indet.       & $ 11.4\pm  0.9$ & $ 11.3\pm  0.4$ & $  0.1\pm  0.4$ \\
 19 & 04:21:06.8 & +26:57:45 &   8.10 & $  2.7\pm  1.2$ & $  4.7\pm  0.6$ & $  0.6\pm  0.2$ & $  3.0\pm  1.2$ & $ 11.1\pm  0.6$ & $ 11.8\pm  0.2$ & $  0.1\pm  0.0$ \\
 20 & 04:21:12.0 & +26:55:51 &   8.75 & $  2.9\pm  0.8$ & $  8.6\pm  0.2$ & $  0.3\pm  0.1$ & $  4.0\pm  1.2$ & $ 10.1\pm  0.7$ & $ 13.9\pm  0.1$ & $ -1.6\pm  0.6$ \\
\hline
\end{tabular}
\end{minipage}
\end{table*}

\begin{table*}
 \centering
 \begin{minipage}{140mm}
  \caption{Comparison of masses and temperatures estimated using the three 
different techniques discussed in the text}
  \begin{tabular}{@{}rccccccccccccc@{}}
  \hline
   & & \multicolumn{2}{c}{SED-fitting\footnote{Based on spatially integrated fluxes.}}
& & & \multicolumn{3}{c}{COREFIT}
& & &\multicolumn{3}{c}{COREFIT-PH} \\
   Core & & Mass & $T$ & & & Mass      & $T_0$& $\bar{T}$\footnote{Density-weighted mean value of $T(r)$ for $r \le r_{\rm out}$.} & & & Mass     & $T_0$ & $\chi_{\rm ISRF}$\footnote{Estimated ISRF scaling factor.} \\
   No. & & [$M_\odot$] & [K] & & & [$M_\odot$]& [K]& [K] & & & [$M_\odot$] & [K] & \\
 \hline
  1 & & $1.19\pm0.05$ & $ 9.9\pm 0.3$ & & & $  0.20 \pm  0.01$ & $   9.5 \pm   0.5$& 10.8 & & &  0.25 &   8.2 & 0.28 \\
  2 & & $0.71\pm0.04$ & $11.8\pm 0.5$ & & & $  0.84 \pm  0.04$ & $  10.7 \pm   0.3$& 11.8 & & &  0.72 &   8.4 & 0.90 \\
  3 & & $0.10\pm0.01$ & $12.3\pm 0.5$ & & & $  0.09 \pm  0.04$ & $  12.3 \pm   0.8$& 12.7 & & &  0.23 &   9.1 & 0.37 \\
  4 & & $0.05\pm0.01$ & $12.6\pm 0.8$ & & & $  0.04 \pm  0.01$ & $   7.1 \pm   0.6$& 9.3  & & &  0.08 &   9.1 & 0.29 \\
  5 & & $0.20\pm0.01$ & $11.3\pm 0.1$ & & & $  0.37 \pm  0.02$ & $   9.4 \pm   0.4$& 10.9 & & &  0.32 &   8.0 & 0.37 \\
  6 & & $1.01\pm0.15$ & $ 8.0\pm 0.6$ & & & $  1.70 \pm  0.34$ & $   7.0 \pm   0.1$& 7.7  & & &  1.64 &   5.0 & 0.26 \\
  7 & & $0.22\pm0.10$ & $ 8.8\pm 1.1$ & & & $  0.49 \pm  0.09$ & $   7.1 \pm   0.4$& 8.6  & & &  0.37 &   6.0 & 0.31 \\
  8 & & $0.07\pm0.01$ & $ 9.9\pm 1.5$ & & & $  0.19 \pm  0.04$ & $   9.9 \pm   1.6$& 10.3 & & &  0.25 &   6.7 & 0.14 \\
  9 & & $1.53\pm0.02$ & $ 9.2\pm 0.1$ & & & $  1.44 \pm  0.09$ & $   8.0 \pm   0.2$& 9.2  & & &  1.53 &   5.5 & 0.39 \\
 10 & & $1.30\pm0.20$ & $ 8.8\pm 0.9$ & & & $  0.96 \pm  0.02$ & $  10.0 \pm   0.6$& 9.9  & & &  1.35 &   5.8 & 0.32 \\
 11 & & $0.20\pm0.03$ & $11.8\pm 1.1$ & & & $  0.32 \pm  0.06$ & $   9.8 \pm   0.6$& 10.6 & & &  0.41 &   7.4 & 0.29 \\
 12 & & $0.42\pm0.05$ & $ 9.2\pm 0.6$ & & & $  0.41 \pm  0.06$ & $   9.9 \pm   1.1$& 10.1 & & &  0.40 &   6.6 & 0.39 \\
 13 & & $1.60\pm0.28$ & $ 8.8\pm 1.0$ & & & $  2.03 \pm  0.16$ & $   6.0 \pm   0.5$& 8.3  & & &  2.21 &   5.4 & 0.32 \\
 14 & & $0.47\pm0.11$ & $ 8.8\pm 1.7$ & & & $  0.46 \pm  0.05$ & $   7.8 \pm   0.4$& 9.4  & & &  0.33 &   6.7 & 0.25 \\
 15 & & $0.21\pm0.07$ & $11.2\pm 1.9$ & & & $  0.84 \pm  0.08$ & $   8.8 \pm   0.4$& 10.0 & & &  1.04 &   6.1 & 0.36 \\
 16 & & $0.36\pm0.06$ & $ 8.7\pm 0.5$ & & & $  1.08 \pm  0.12$ & $   6.4 \pm   0.2$& 8.3  & & &  1.52 &   5.0 & 0.26 \\
 17 & & $0.05\pm0.01$ & $12.5\pm 1.2$ & & & $  0.08 \pm  0.01$ & $  10.7 \pm   0.4$& 11.9 & & &  0.12 &   8.8 & 0.18 \\
 18 & & $0.21\pm0.03$ & $ 9.6\pm 0.8$ & & & $  0.09 \pm  0.02$ & $  11.4 \pm   0.9$& 11.3 & & &  0.07 &   9.2 & 0.25 \\
 19 & & $0.08\pm0.01$ & $11.8\pm 0.8$ & & & $  0.08 \pm  0.02$ & $  11.1 \pm   0.6$&11.5  & & &  0.09 &   8.6 & 0.38 \\
 20 & & $0.02\pm0.01$ & $13.6\pm 0.7$ & & & $  0.06 \pm  0.01$ & $  10.1 \pm   0.7$&11.8  & & &  0.05 &  10.2 & 0.28 \\
\hline
\end{tabular}
\end{minipage}
\end{table*}

\section[]{Comparison with previous observations}

The deduced physical properties of our cores may be compared with previously
published spectral line data in H$^{13}$CO$^+$ and N$_2$H$^+$, both of
which are known to be good tracers of high density gas ($n({\rm H}_2)
\stackrel{>}{_\sim} 10^5$ cm$^{-3}$).  Of our 20 cores, we find
accompanying observations for 10 in
H$^{13}$CO$^+$ \citep{onishi2002} and seven in N$_2$H$^+$ 
\citep{hac2013}.  The relevant parameters are given in Table 4. 

\begin{table*}
 \begin{minipage}{170mm}
  \caption{Comparison with previously published spectral line data.}
  \begin{tabular}{@{}cccccccccccccccc@{}}
  \hline
Core\footnote{As listed in Table 2.} 

& \multicolumn{2}{c}{ID\footnote{Object number in previously published 
source lists: \citet{onishi2002} in H$^{13}$CO$^+$,
and \citet{hac2013} in N$_2$H$^+$.}} &

& \multicolumn{2}{c}{Offset\footnote{Angular offset from the COREFIT (dust continuum) position.} [arcsec]} &

& \multicolumn{3}{c}{Radius [pc]} &

& \multicolumn{2}{c}{Mass [$M_\odot$]\footnote{The mass quoted in the ``Dust" column represents the COREFIT estimate of total mass (gas + dust) based on the dust thermal continuum in the 70--500 $\mu$m range; the mass quoted for H$^{13}$CO$^+$ represents a virial mass derived by \citet{onishi2002}.}} &

& \multicolumn{2}{c}{$n$(H$_2$) [$10^5$ cm$^{-3}$]} \\

No. &Onishi&Hacar& & H$^{13}$CO$^+$ & N$_2$H$^+$ & & Dust\footnote{The radius quoted here is $r_{\rm out}$ from Table 2, converted to pc.} & H$^{13}$CO$^+$ & N$_2$H$^+$ & & Dust & H$^{13}$CO$^+$ 
& & Dust & H$^{13}$CO$^+$ \\
 \hline
1 & 3 & \ldots & & 41 &\ldots& & 0.034 & 0.021 &\ldots&  & 0.2& 1.7 & & 0.5 & 0.9 \\
4 & 5 & \ldots & & 74 &\ldots& & 0.041 & 0.054 &\ldots&  & 0.04& 6.5 & & 9.8 & 1.9 \\
6 & 5 &   1    & & 93 &  43  & & 0.043 & 0.054 & 0.048&  & 1.7& 6.5 & & 25 & 1.9 \\
7 & 5 & \ldots & & 91 &\ldots& & 0.034 & 0.054 &\ldots&  & 0.5& 6.5 & & 13 & 1.9 \\
9 &\ldots& 2   & &\ldots&8.4 & & 0.050 &\ldots & 0.027&  & 1.4&\ldots& &1.9& \ldots \\
10& 6 & \ldots & & 52 & \ldots & & 0.054 & 0.034 & 0.051&  & 1.0& 5.8 & & 0.9 & 1.2 \\
12& 7 &    5   & & 82 & 31   & & 0.042 & 0.035 & 0.030&  & 0.4& 2.9 & & 2.3 & 1.9 \\
13& 8 &    6   & & 24 & 44   & & 0.065 & 0.064 & 0.053&  & 2.0& 5.0 & & 6.8 & 1.0 \\
14& 9 &    7   & & 44 & 21    & & 0.035 & 0.060 & \ldots &  & 0.5& 4.2 & & 1.9 & 1.0 \\
15& 13a & 10   & & 8.0& 19   & & 0.052 & 0.048 & 0.047&  & 0.8& 3.4 & & 1.6 & 1.4 \\
16&\ldots& 12  & & \ldots&9.4& & 0.043 & \ldots& 0.034&  & 1.1&\ldots& &3.9 &\ldots \\
18& 16a&\ldots & & 59 &\ldots& & 0.025 & 0.028 &\ldots&  & 0.09& 3.0 & & 0.3 & 2.5 \\
\hline
\end{tabular}
\end{minipage}
\end{table*}

Considering first the H$^{13}$CO$^+$ data, comparison of observed peak 
locations with dust continuum source positions from COREFIT shows a distinct 
lack of detailed correspondence.  This behaviour is apparent in Fig. \ref{fig1}
and from Table 4 which includes
the angular distance (labeled as ``Offset" in the table) between each of the H$^{13}$CO$^+$ source locations
and the corresponding dust continuum core location.
The median distance is $59''$, considerably larger than the spatial
resolution of either the H$^{13}$CO$^+$ observations ($20''$) or the
{\it Herschel\/} data ($18''$ at 250 $\mu$m).  These offsets are somewhat surprising,
since previous comparisons between H$^{13}$CO$^+$ and dust continuum maps
have shown good correspondence \citep{um02}. One could question whether
they are due to gridding errors in the H$^{13}$CO$^+$ data,
in view of the fact that the observations were made
on a relatively coarse grid (the eight cores of \citet{onishi2002}
in Table 4 are split evenly between $30''$ and $60''$ grid spacings).
However, the measured offsets show no correlation with
the grid spacing---the mean offset is approximately $50''$ in either 
case; this argues against gridding error as an explanation.
The most likely reason for the systematic offsets
is that the H$^{13}$CO$^+$ is frozen out at the low ($\stackrel{<}{_\sim}10$ K)
temperatures of the core centres \citep{walm04}. 

Detailed comparison of the dust continuum core locations with the
H$^{13}$CO$^+$ maps (four examples of which are given in Fig. \ref{fig10}) 
shows that the majority of
sources are elongated and/or double and that in some cases
(Onishi core No. 3 in particular) the dust continuum source falls
between the pair of H$^{13}$CO$^+$ components. In other cases (e.g.
Onishi core No. 16a), the dust continuum peak falls on a 
nearby secondary maximum of the H$^{13}$CO$^+$ emission. In the latter case, 
surprisingly, the main peak of H$^{13}$CO$^+$ falls in a local {\it minimum\/} 
of dust emission.  Comparisons between H$^{13}$CO$^+$ images and their
250 $\mu$m counterparts show that, in general, the elongation and source
alignment in H$^{13}$CO$^+$ is along the filament, so we have a rod-like,
rather than spherical, geometry. The picture which thus
emerges is that when a core forms in a filament \citep{andre10},
we see the core centre in dust continuum emission and the warmer
(but still dense, $\sim10^5$ cm$^{-3}$) H$^{13}$CO$^+$ gas on either side of it
in a dumbbell-like configuration.  The median separation of the dust continuum 
and H$^{13}$CO$^+$ sources then corresponds to the typical radius of the 
depletion zone.  For an ensemble of randomly oriented filaments, the
mean projected separation is $\pi/4$ times the actual separation, which
means that our estimated median separation of $59''$ corresponds to
an actual separation of $75''$, or about $1.1\times10^4$ AU at the distance
of L1495.  This is similar to the radius of the dark-cloud chemistry zone
in which carbon-bearing molecules become gaseous \citep{cas11}.

\begin{figure}
\includegraphics[width=88mm]{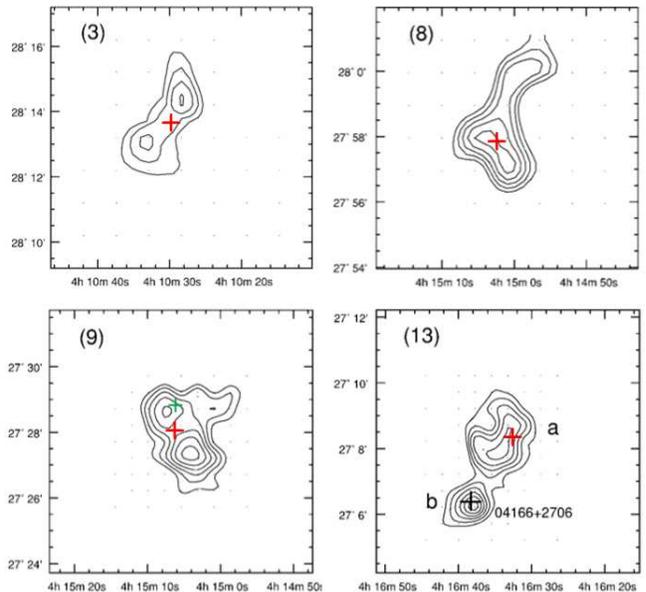}
 \caption{Examples of dust continuum core locations in relation to
H$^{13}$CO$^+$ emission.  The estimated locations of cores 1, 13, 14 and 15 
(corresponding to Onishi core Nos. 3, 8, 9, and 13a, respectively)
are superposed on H$^{13}$CO$^+$ maps taken from \citet{onishi2002}
(B1950 coordinates).
In each case, the location of peak dust column density is indicated by a red 
cross. The green cross in Onishi field (9) represents a secondary peak of dust 
emission. The black cross in Onishi field (13) represents a protostar location.}
 \label{fig11}
\end{figure}

Comparing the estimated masses, Table 4 shows that the values estimated 
from dust continuum observations are, in most cases, much smaller than
those based on H$^{13}$CO$^+$. The discrepancy ranges from a factor of
$\sim2$ to more than an order of magnitude. 
Based on the mass and positional discrepancies
it is clear that H$^{13}$CO$^+$ and
submillimetre continuum are not mapping the same structures.  
Nevertheless, it remains to explain why so much of the expected dust emission
from the H$^{13}$CO$^+$ emitting gas is apparently not being seen in the submillimetre
continuum.  It is unlikely to be a result of the background subtraction
in COREFIT since
the COREFIT mass estimates match the SED-based values from {\it getsources\/} 
fluxes within $\sim30$\% and the only background that was subtracted during 
the latter processing corresponded to the natural
spatial scale of the broader underlying emission structure.

The most likely explanation for the discrepancy is an overestimation of
the virial mass of the gas component due, in part, to the assumption by
\citet{onishi2002} of uniform velocity dispersion.
Specifically, the velocity dispersion of the relatively cool
gas being probed by dust emission
is likely to be at least a factor of two lower than that of 
H$^{13}$CO$^+$, as suggested by the
N$_2$H$^+$ observations of \citet{hac2013},
and since the estimated virial mass depends on the square of that dispersion,
it could have been overestimated
by a factor of up to 4.  Two additional effects, both of which are likely
to have led to overestimation of the virial mass are:
\begin{enumerate}
  \item the \citet{onishi2002} virial mass was 
based on assumed spherical shape as opposed to the filamentary
geometry observed, and hence the source volumes may have been
overestimated;
  \item \citet{onishi2002} assumed a constant density value for each core.
However, virial models involving this assumption are likely to lead to
overestimates of mass in cases where the actual density decreases outwards
\citep{mac88}.
\end{enumerate}

While the H$^{13}$CO$^+$ peaks do not correlate well with the dust continuum,
the situation is different for N$_2$H$^+$. This behaviour can be seen from 
Table 4
which includes the positional offsets between N$_2$H$^+$ and dust continuum
peaks; the median offset is only $21''$, i.e., only a third of
the corresponding value for H$^{13}$CO$^+$ even though the resolution of
the N$_2$H$^+$ observations ($60''$) was much coarser.  Thus the positional
data provide no evidence for N$_2$H$^+$ freeze-out, and this is supported
by the fact that the N$_2$H$^+$ detections seem preferentially associated
with the coldest cores (the seven N$_2$H$^+$ detections include four of our 
five lowest-temperature cores, all of which are cooler than 7 K). 
However, at higher resolution the situation
may be different, since interferometric observations of
$\rho$ Oph have shown that the correspondence between dust emission and
N$_2$H$^+$ clumps does indeed break down on spatial scales 
$\stackrel{<}{_\sim}10''$ \citep{fries2010}.  
Theoretical models have, in fact, shown that within
$\sim1000$ AU of the core centre, complete freeze-out of heavy elements
is likely \citep{cas11}.  Core profiling based on dust emission thus promises
to make an important contribution to investigations of core chemistry by 
providing an independent method for estimating temperatures in the centres 
of cores.

Finally, our core No. 16 has been observed previously in the 850 $\mu$m
continuum by \citet{sad2010} and given the designation JCMTSF\_041950.8+271130.
While the quoted 850 $\mu$m source radius of 0.019 pc is close to the
COREFIT $r_0$ value of $0.015\pm0.002$ pc, the estimated masses are 
significantly different. 
The estimate of \citet{sad2010} is based on the observed 850 $\mu$m flux
density and an assumed dust temperature of 13 K; this yielded 0.22 $M_\odot$ 
which is a factor $\sim5$ smaller than our COREFIT value and most likely
an underestimate resulting from too high an assumed temperature.
This illustrates the large errors in mass which can occur in the absence of 
temperature information, as has been noted by others \citep{stam07,hill2009,hill2010}.  

\section[]{Conclusions}
The principal conclusions from this study can be summarised as follows:
\begin{enumerate}
  \item For this sample of cores, the dust temperatures estimated
from SED fits, using spatially integrated fluxes and an isothermal model, are 
consistent with the spatially averaged temperatures derived from the COREFIT 
profiles. However, the masses 
obtained from the SED fits are systematically
lower (by a factor of $\sim1.5$) than those obtained from detailed core 
profiling.  The present statistical sample, however, is insufficient to obtain
a definitive correction factor, the latter of which is likely to be
dependent on mass and possibly environment (ISRF) also.
  \item Estimates of central dust temperature are in the range 6--12 K. 
These temperatures are negatively correlated with peak column density, 
consistent with behaviour expected
due to shielding of core centre from the ISRF, assuming that the latter
provides the sole heating mechanism.
The model core temperatures obtained from radiative transfer calculations
are, however, systematically $\sim2$ K lower than the COREFIT estimates; it
is not yet clear whether that difference has an astrophysical origin or
is due to errors in model assumptions.
  \item The radial falloff in density is, in the majority of cases, 
consistent with the $\alpha\simeq2$ variation expected for 
Bonnor-Ebert spheres although exceptions are found in two cases, both
of which appear to have steeper density profiles. Since both involve
cores which are gravitationally unstable based on Bonnor-Ebert criteria, such
behaviour may have implications for models of the early collapse phase.
  \item The reliability of core stability estimates derived from isothermal
models is not seriously impacted by the temperature gradients 
known to be present in cores. Thus the preliminary classification of 
cores as gravitationally bound or unbound can be based on relatively simple 
criteria, facilitating statistical studies of large samples.
 \item Core locations do not correspond well with previously published locations
of H$^{13}$CO$^+$ peaks, presumably because carbon-bearing molecules are
frozen out in the central regions of the cores, most of which have dust
temperatures below 10 K. The results suggest that the H$^{13}$CO$^+$ emission
arises from dense gas in the filamentary region on either side of the
core itself, in a dumbbell-like geometry, and that the radius of the 
sublimation zone is typically $\sim10^4$ AU. 
 \item The coldest cores are mostly detected in N$_2$H$^+$, and the 
N$_2$H$^+$ core locations are consistent with those inferred from
dust emission, albeit at the relatively coarse ($1'$) resolution of
the N$_2$H$^+$ data.  Our data therefore do not show evidence of
N$_2$H$^+$ freeze-out.
\end{enumerate}

\section*{Acknowledgments}

We thank T. Velusamy, D. Li, and P. Goldsmith for many helpful discussions
during the early development of the COREFIT algorithm. We also thank the
reviewer for many helpful comments.
SPIRE has been developed by a consortium of institutes led by
Cardiff Univ. (UK) and including: Univ. Lethbridge (Canada);
NAOC (China); CEA, LAM (France); IFSI, Univ. Padua (Italy);
IAC (Spain); Stockholm Observatory (Sweden); Imperial College
London, RAL, UCL-MSSL, UKATC, Univ. Sussex (UK); and
Caltech, JPL, NHSC, Univ. Colorado (USA). This development
has been supported by national funding agencies: CSA (Canada);
NAOC (China); CEA, CNES, CNRS (France); ASI (Italy); MCINN
(Spain); SNSB (Sweden); STFC, UKSA (UK); and NASA (USA).
PACS has been developed by a consortium of institutes led by
MPE (Germany) and including UVIE (Austria); KU Leuven, CSL,
IMEC (Belgium); CEA, LAM (France); MPIA (Germany); INAFIFSI/
OAA/OAP/OAT, LENS, SISSA (Italy); IAC (Spain). This development
has been supported by the funding agencies BMVIT
(Austria), ESA-PRODEX (Belgium), CEA/CNES (France), DLR
(Germany), ASI/INAF (Italy), and CICYT/MCYT (Spain).

\bsp

\label{lastpage}

\end{document}